\documentclass[journal]{IEEEtran}
\IEEEoverridecommandlockouts

\usepackage{relsize}
\usepackage{moreverb}
\usepackage{mathrsfs}
\usepackage{amsmath}
\usepackage{amssymb}
\usepackage{stfloats}
\usepackage{graphicx}
\usepackage{setspace}
\usepackage{xcolor}
\usepackage{cite}
\usepackage{algorithm}      
\usepackage{algorithmic}    
\usepackage{multirow}       

\usepackage{tabularx,ragged2e,booktabs,caption}

\ifCLASSOPTIONcompsoc
\usepackage[tight,normalsize,sf,SF]{subfigure}
\else
\usepackage[tight,footnotesize]{subfigure}
\fi

\usepackage{amsmath}

\begin{document}
%

%
\title{Deep Learning Assisted Antenna Selection in Untrusted Relay Networks}

\author{
\IEEEauthorblockN{Rugui Yao, Yuxin Zhang, Shengyao Wang, Nan Qi, Theodoros A. Tsiftsis and Nikolaos I. Miridakis}
}

\maketitle

\begin{abstract}
This letter mainly studies the \emph{transmit antenna selection}(TAS) based on \emph{deep learning} (DL) scheme in untrusted relay networks.
In previous work, we discover that \emph{machine learning} (ML)-based antenna selection schemes have small performance degradation caused by complicated coupling relationship between achievable secrecy rate and the channel gains.
To solve the issue, we here introduce \emph{deep neural network} (DNN) to decouple the complicated relationship.
The simulation results show the DNN scheme can achieve better decoupling and thus perform almost the same performance with conventional exhausted searching scheme.
\end{abstract}


\IEEEpeerreviewmaketitle

\section{introduction}
In recent years, \emph{artificial intelligence} (AI) has made great success in many fields such as pattern recognitions and signal processing, and the intelligence communication is considered a sightful way for wireless communication after 5G \cite{Li2017Intelligent}.
Currently, the research of AI is advancing into physical layer security in wireless communication so as to reduce the complexity and solve other existing problems to improve the system performance \cite{D2017Deep}.

Especially, the ML technology has become one of most popular AI technologies in physical layer security \cite{Oshea2017An}.
For example, the ML schemes are conducted for the antenna selection in wiretap networks \cite{He2018Transmit}. In \cite{8419776}, the author studied resource allocation in multi-channel cognitive networks using DNN method. DL algorithm was proposed in \cite{7852251} to improve the performance of the belief propagation algorithm for decoding. In \cite{O2017Deep}, an unsupervised deep learning was used in \emph{Multiple-Input Multiple-Output} (MIMO) encoder system.

In our previous work \cite{icc_yao}, we applied ML schemes, namely, \emph{support vector machine}(SVM), \emph{naive-bayes}(NB), and \emph{k-nearest neighbors}(k-NN), to implement TAS for the optimization problem in (\ref{opt_prob}). However, ML-based scheme has a little system degradation due to nonlinear coupling. In this paper, since the DNN can solve the various nonlinear distortions, we believe that the DL application is a useful and promising way to achieve decoupling in antenna selection. Our main contributions are as follows:
\begin{itemize}
  \item We focus on DL scheme to enhance physical layer security in untrusted relay networks.
  \item Compared to our previous work in  \cite{icc_yao}, we extend our research by DNN to achieve decoupling and further get better system performance.
\end{itemize}

\section{System and Signal Model}\label{signal Model}
We consider half-duplex two-hop untrusted relay networks, consisting of a source (S), a destination (D) and an untrusted \emph{amplifying-and-forward} (AF) relay (R) equipped with $N_\mathrm{S}$, $N_\mathrm{R}$, $N_\mathrm{D}$ antennas, respectively.
For simplicity, we assume $N_\mathrm{R} = N_\mathrm{D} =1$ for our initial work. We note that only the source S is employed to process the transmit antenna selection. Further, all channels are subject to \emph{independent and identically distributed} (i.i.d) Rayleigh fading.


In this system, there is no direct link between S and D because of shadowing or long distance; Therefore the communication is implemented via R.
We denote $\mathbf{h} = \left[h_1, \cdots, h_{N_\mathrm{S}}\right] \in \mathcal{C}^{1 \times N_\mathrm{S}}$ as the channel vector from S to R.
Further define $g_\mathrm{R-D} \in \mathcal{C}^{1 \times 1}$ and $g_\mathrm{D-R} \in \mathcal{C}^{1 \times 1}$  as the channel gains from R to D and from D to R. Here, since the channel reciprocity is considered, we have $g_\mathrm{R-D} = g_\mathrm{D-R}^*$ = $g$.
Due to the high cost of RF chain, only $N_\mathrm{T}$ antennas among $N_\mathrm{S}$ of S are activated to perform transmission.
Assume that the available $N_\mathrm{S}$ antennas are labeled as $1, 2, \cdots, N_\mathrm{S}$ and the selected $N_\mathrm{T}$ antennas are with the indices $s_1,s_2, \cdots, s_{N_\mathrm{T}}$ where $s_j \in [1, N_\mathrm{S}]$ for $j=1, \cdots, N_\mathrm{T}$. Therefore, the practical propagation channel from S to R can be denoted as $\widetilde{\mathbf{h}}=\left[h_{s_1}, \cdots, h_{s_{N_\mathrm{T}}}\right] \in \mathcal{C}^{1 \times N_\mathrm{T}}$.
In order to maximize the received SNR for the relay R, the source S adopts matched filter precoding. In this case, the precoding vector for S's transmission is $\mathbf{p}_\mathrm{MF} = \frac{\widetilde{\mathbf{h}}^{\mathrm{H}}}{\|\widetilde{\mathbf{h}}\|_2}$, where $\|\cdot\|_2$ represents the $2$-norm of a vector.

Owing to the relay is untrustworthy, we adopt the \emph{destination-aided jamming} (DAJ) technique\cite{yao_access} and divide the transmission into two time-slots.

In the first time slot, S transmits its precoded signal, $\mathbf{p}_\mathrm{MF} x_\mathrm{S}$ to R with $x_S$ being the confidential signal, and simultaneously D emits cooperative jamming signal $x_\mathrm{J}$ to R in the same frequency. The received signal at R, $y_\mathrm{R}$, is presented by \cite{yao_access}

\begin{align}\label{y_R}
y_\mathrm{R} &= \sqrt{\frac{P_\mathrm{S}}{N_\mathrm{T}}} \widetilde{\mathbf{h}} \mathbf{p}_\mathrm{MF} x_\mathrm{S} + \sqrt{P_\mathrm{D}} g_\mathrm{D-R}x_\mathrm{J} + n_\mathrm{R}\nonumber\\
&=\sqrt{\frac{P_\mathrm{S}}{N_\mathrm{T}}} \|\widetilde{\mathbf{h}}\|_2 x_\mathrm{S} + \sqrt{P_\mathrm{D}} g x_\mathrm{J} + n_\mathrm{R},
\end{align}
\noindent where $x_\mathrm{S}$ and $x_\mathrm{J}$ are both with unit power;
$P_\mathrm{S}$ and $P_\mathrm{D}$ are the transmitted powers from $S$ and $D$;
$n_\mathrm{R}$ denotes the complex \emph{additive white Gaussian noise} (AWGN) received at R, following $\mathcal{CN}(0, N_0)$-distribution.
In this letter, all the AWGNs received both at R in the first time slot and at D in the second time slot are assumed with unit \emph{power spectral density} (PSD), i.e., $N_0 = 1$. Hence, the \emph{signal-to-noise ratio} (SNR) at different nodes can be adjusted by the transmitted power. With the cooperative jamming signal as the second item in (\ref{y_R}), the eavesdropping capability of R is degraded.

From (\ref{y_R}), the instantaneous received \emph{signal-to-interference-plus-noise ratio} (SINR) at R can be denoted as
\begin{align}\label{gamma_R}
\gamma_\mathrm{R} = \frac{\frac{P_\mathrm{S}}{N_\mathrm{T}} \|\widetilde{\mathbf{h}}\|_2^2
}
{\left(P_\mathrm{D} |g|^2 + 1\right)}.
\end{align}

In the second time slot, the relay R re-transmits the received signal to D after amplifying it with an amplification factor $\beta$. Let $P_\mathrm{R}$ be the transmitted power by R. Therefore, with $y_\mathrm{R}$ in (\ref{y_R}), the amplification factor $\beta$ can be denoted as
\begin{align}\label{beta2}
\beta^2 = \frac{P_\mathrm{R}}{\frac{P_\mathrm{S}}{N_\mathrm{T}}
\|\widetilde{\mathbf{h}}\|_2^2
+ P_\mathrm{D} |g|^2 + 1}.
\end{align}

Then the received signal at D from R is given by

\begin{align}\label{y_D}
y_\mathrm{D} &= \beta g_\mathrm{R-D} \frac{\sqrt{P_\mathrm{S}}}{N_\mathrm{T}}
\|\widetilde{\mathbf{h}}\|_2^2
x_\mathrm{S} + \beta g_\mathrm{R-D} \sqrt{P_\mathrm{D}} g_\mathrm{D-R}x_\mathrm{J} \nonumber\\
&~~~+ \beta g_\mathrm{R-D} n_\mathrm{R} + n_\mathrm{D},
\end{align}
\noindent where $n_\mathrm{D}$ is the complex AWGN received at D, which is also assumed to be $\mathcal{CN}(0, 1)$-distributed.

Since the second item in (\ref{y_D}) is transmitted by D itself, D can perform self-interference cancellation with perfect \emph{channel state information} (CSI) available.
Consequently, the instantaneous SINR at D can be presented as
\begin{align}\label{gamma_D}
\gamma _\mathrm{D}=\frac{\frac{P_\mathrm{S}}{N_\mathrm{T}} \beta^2 \| \widetilde{\mathbf{h}}\|_2^2 | g |^2}{\beta^2 | g |^2 + 1}.
\end{align}

In physical-layer security based untrusted relay system, the achievable secrecy rate can be defined as \cite{yao_tifs}
\begin{align}\label{R_s}
R_s=\left[\log_2(1+\gamma_\mathrm{D})-\log_2(1+\gamma_\mathrm{R})\right]^+,
\end{align}
\noindent where $[\cdot]^+=\max(\cdot, 0)$. Note that, for simplification, we neglect this operator for the following derivation but consider it for simulation.

With (\ref{gamma_R}) and (\ref{gamma_D}), we can formulate the secrecy rate $R_s$ in (\ref{R_s}) after some calculations as
\begin{align}\label{R_s_all}
R_s&=\log_2 \left(
\frac{\frac{P_\mathrm{S}}{N_{\mathrm{T}}} \|\widetilde{\mathbf{h}}\|_2^2}{P_\mathrm{D}|g|^2+1}
\right.\nonumber\\
&\times \left.
\frac{\frac{P_\mathrm{S}}{N_{\mathrm{T}}} P_\mathrm{R} |g|^2 \|\widetilde{\mathbf{h}}\|_2^2
+P_\mathrm{R} |g|^2+\frac{P_\mathrm{S}}{N_\mathrm{T}}
\|\widetilde{\mathbf{h}}\|_2^2
+ P_\mathrm{D} |g|^2 + 1
}
{P_\mathrm{R}|g|^2+\frac{P_\mathrm{S}}{N_\mathrm{T}}
\|\widetilde{\mathbf{h}}\|_2^2
+ P_\mathrm{D} |g|^2 + 1}
\right).
\end{align}

When we select only one antenna, namely $h_s$, to implement the transmission. Therefore, $\|\widetilde{\mathbf{h}}\|^2_2$ in (\ref{R_s_all}) can be further replaced with $|h_s|^2$.

\section{Conventional And Machine Learning-based Antenna Selection Scheme}\label{con sch}
In conventional antenna selection, the source S can be aware of all CSIs, such as $\mathbf{h}$ and $g$. Then, S traverses all the possible antenna combinations, and computes the corresponding secrecy rate. The maximum secrecy rate and the corresponding antenna selection scheme are the solutions for TAS problem. The optimization problem can be formulated as
\begin{align}\label{opt_prob}
n^* = \arg \max_{n \in \mathcal{L}} R_s,
\end{align}
\noindent where $\mathcal{L}$ denote the index set for all possible combinations for selected antennas, with size $
\mathbb{C}_{N_\mathrm{S}}^{N_\mathrm{T}} =\frac{N_\mathrm{S}!}{N_\mathrm{T}!(N_\mathrm{S}-N_\mathrm{T})!}$.

As can be seen from the analysis of (\ref{R_s_all}), there exists quite an involved coupling relationship for $R_s$ with $\widetilde{\mathbf{h}}$ and $g$. It is hard for ML scheme to achieve decoupling since the ML schemes usually deal with some linear problems; in this case more misclassification is led to and degradation of system performance is emerged \cite{icc_yao}. Considering the superiority of DNN to solve nonlinearity, we will introduce DNN method to decouple the above complicated relationship and achieve the optimal antenna selection in untrusted relay networks.

\section{DNN-based antenna selection scheme}
The DNN structure we utilize here has 3 layers consisting of an input layer, hidden layers and an output layer and every layer has their own neurons \cite{Liu2017A}. Three procedures for DNN-based TAS scheme are as follows.
\subsection{Data sets generation}
First, we generate a training data set and a testing data set, each of which containing $M$ diverse real $1 \times (N_S+1)$ \emph{channel state information}(CSI) data samples, e.g. $\mathbf{d}^1,\mathbf{d}^2, \cdots,\mathbf{d}^M$. To be specific, the $m$-th training or testing data samples can be denoted as $\mathbf{d}^m = \left[|h^m_1|, |h^m_2|, \cdots, |h^m_{N_\mathrm{S}}|, |g^m|\right]$£¬ for $m\in \{1,\cdots, M\}$.

Then, we generate normalized feature vector $\mathbf{t}^{m}$ by normalizing $\mathbf{d}^{m}$. The $i$-th element of $\mathbf{t}^{m}$, $t_i^m$, can be generated as
      \begin{align}\label{normalize}
       t_i^m = \frac{d_i^m -  \mathbb{E}[\mathbf{d}^m]}{\max(\mathbf{d}^m) -  \min(\mathbf{d}^m)},
      \end{align}
      \noindent
where $d_i^m$ is the $i$-th element of $\mathbf{d}^{m}$; $\mathbb{E}$ denotes the expected value operation.

Furthermore, we calculate secrecy rate of each antenna combination in $\mathcal{L}$ as the KPI, and we choose the target antennas that achieves the maximum secrecy rate.
\subsection{Construct DNN model}
We construct DNN model on Tensorflow platform.

In input layer of the training DNN model, each item of the normalized training feature vector obtained in (\ref{normalize}) corresponds to each neuron as the input.

In output layer of the DNN model, selected labels corresponds to each neuron as the output. Meanwhile, The one-hot encoding is choosed for the labels. It means that when there is $\mathcal{L}$ combinations, it needs $\mathcal{L}$ bits to encode labels. The processed label has only one bit equaling to 1 and other bits equaling to 0. For example, when selecting 1 antenna from 6 antennas, the selected label 6 can be coded as 000001 and 4 can be coded as 000100; when selecting 2 antenna from 6 antennas, the selected label 10 can be coded as 000000000100000 and 14 can be coded as 000000000000010.

Then, we set a series of parameters, such as learning rate, model training times, batch size and the neurons of input layer, hidden layer and output layer and so on. We adopt \emph{rectified linear units} (ReLU) function as the hidden layer function; In addition, we apply RMSProp optimizer as the model optimizer method. For last layer's neuron $x$ to next layer's neuron $y$, it can be denoted as $y={\bf{W}}^{\rm{T}}  x+b$, The weight matrix of last layer's output to the next layer's neuron is $\bf{W}$; $b$ is the bias parameter.
In addition, the RMSProp optimizer as the model optimizer method is applied.

\subsection{Start model training and predication}
We train DNN model to extract its features and set up internal parameters. After the model is established, we perform label prediction of normalized testing feature vector by DNN.

When DNN model carries out label predicting, the normalized testing feature vector obtained in (\ref{normalize}) is regarded as input, and the probability of each antenna (i.e., $\mathcal P_1, \mathcal P_2, \cdots, \mathcal P_{N_s}$) corresponding each output layer's neuron is regarded as output. For example, when it is single antenna selection, the label which make the highest probability will be selected; when it is two antenna selection, the label combination which make the highest probability will be selected.

The reason for setting up probability of each antenna is that we use logistic function equation $f(c) = \frac{1}{{1{\rm{ + }}{e^{ - c}}}}$ as the output layer function, where $c$ correspond the input elements of the output layer.

\section{Simulation Results And Analysis}\label{simulation}
In this section, we present some simulation results to verify the efficiency of the DNN-based schemes.
The size of both training data and testing data are set 200000$\times$7.
The source S is configured with $N_\mathrm{S} = 6$ antennas, and $N_\mathrm{T} = 1$ or $2$ antennas will be selected out. For simplicity, we set $P_\mathrm{S} =P_\mathrm{D} =P_\mathrm{R}$.

The experiment is carried out in Tensorflow platform which exploiting GPU processing power, thus solving the large amount of data more efficiently. By looking for the maximum classification accuracy as possible, those parameters are adjusted and confirmed. The amount of the input layer, hidden layer, output layer of DNN architecture is set to 1, 1, 1, repectively; the batch size is 128; the learning rate is 0.01; when selecting 1 antenna from 6 antennas, the neurons of input layer, hidden layer and output layer are set 7, 1500, 6 or set 7, 1500, 15 when selecting 2 antenna from 6 antennas.


\subsection{System Performance}\label{SP}
The system performance simulation is conducted in terms of secrecy rate defined in (\ref{R_s}) and \emph{secrecy outage probability} (SOP). The SOP can be defined as

\begin{align}\label{SOP}
P_\mathrm{out}(R_t) = \mathbb{P}(R_s < R_t),
\end{align}

\noindent where $\mathbb{P}(\cdot)$ is the probability, $R_t$ is the target SOP and $R_t=2 \rm bps/Hz $.

%



\begin{figure}[ht]
    \centering
    \includegraphics[width=0.45\textwidth] {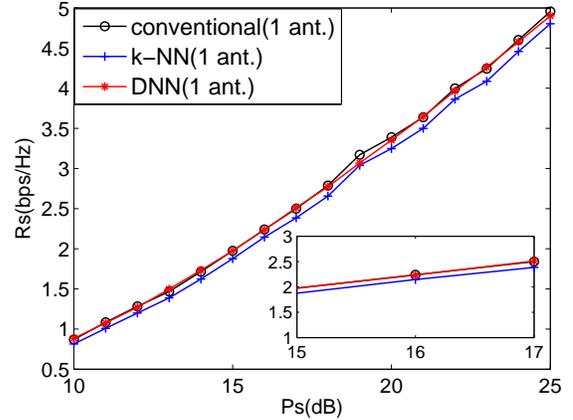}
    \caption{Secrecy rate of single antenna selection for DNN and other schemes}\label{DNN_rate}
\end{figure}

\begin{figure}[ht]
    \centering
    \includegraphics[width=0.45\textwidth] {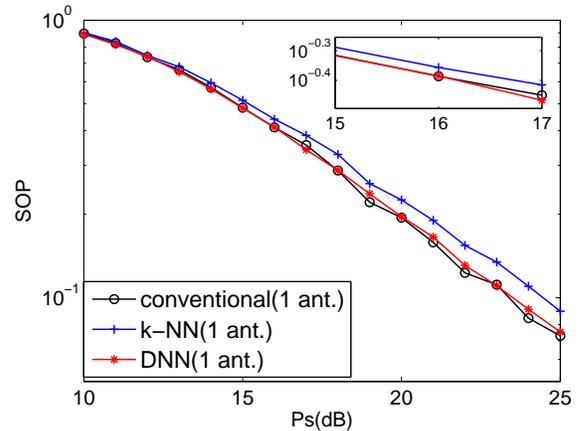}
    \caption{SOP of single antenna selection for DNN and other schemes}\label{DNN_SOP}
\end{figure}

\begin{figure}[ht]
    \centering
    \includegraphics[width=0.45\textwidth] {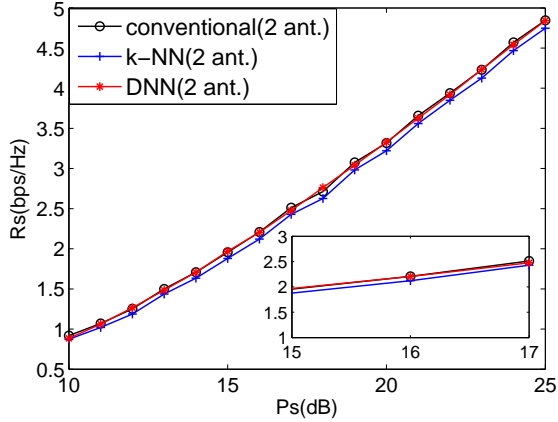}
    \caption{Secrecy rate of two antenna selection for DNN and other schemes}\label{DNN_rate_2}
\end{figure}

\begin{figure}[ht]
    \centering
    \includegraphics[width=0.45\textwidth] {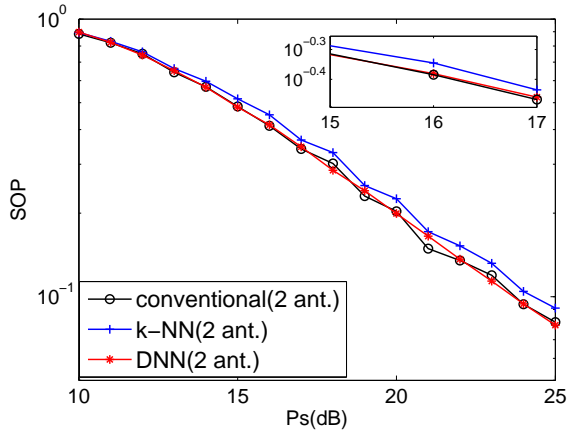}
    \caption{SOP of two antenna selection for DNN and other schemes}\label{DNN_SOP_2}
\end{figure}

From \cite{icc_yao}, the k-NN scheme of ML schemes has the highest performance. Therefore, the single antenna for TAS of the k-NN, DNN, and conventional schemes are compared in terms of secrecy rate and SOP as shown in Figs. \ref{DNN_rate} and \ref{DNN_SOP}. In addition, two antennas for TAS of the k-NN, DNN, and conventional schemes are compared in Figs. \ref{DNN_rate_2} and \ref{DNN_SOP_2}.
And it can indicate that DNN-based TAS scheme outperforms than other ML schemes. Furthermore, the DNN-based scheme achieves almost the same performance with conventional scheme, thus indicating that our proposed DNN-based scheme can achieve decoupling.
It is because that each layer of DNN architecture plays a processing unit to solve the the non-linear relationship among the features while the ML schemes merely solve linear relationship. In addition, the DNN shows outstanding modeling capability compared to existing ML methods.\cite{Liu2017A}

\subsection{Computational Complexity}
As stated in Section \ref{con sch}, $|\mathcal{L}|$ presents the cardinality of selected antenna combinations, and $\left| \mathcal{L} \right| = \mathbb{C}_{N_\mathrm{S}}^{N_{\mathrm{T}}} = \frac{N_{\mathrm{S}}!}{N_{\mathrm{T}}! (N_{\mathrm{S}} - N_{\mathrm{T}})!} $. Let $N = N_\mathrm{S} + 1$.
The selection complexity for SVM, NB, k-NN, DNN and conventional schemes are $\mathcal{O}(N^2)$, $\mathcal{O}(|\mathcal{L}|N+|\mathcal{L}|\log|\mathcal{L}|)$, $\mathcal{O}(N)$, $\mathcal{O}(1)$ and $\mathcal{O}(N+|\mathcal{L}|\log|\mathcal{L}|)$, respectively \cite{Joung2016Machine, He2018Transmit}. We can clearly see that the complexity of DNN and ML schemes are rather lower than that of conventional schemes. It is because that the conventional TAS scheme requires to process the global search and comparison for every antenna combination. However, the complexity of ML-based and DL-based schemes rely on the prediction complexity rather than the training complexity because the model training can be performed offline. Besides, the well-trained DNN architecture only make finite computing processing.

\subsection{Classification Performance}
\begin{figure}[ht]
    \centering
    \includegraphics[width=0.23\textwidth] {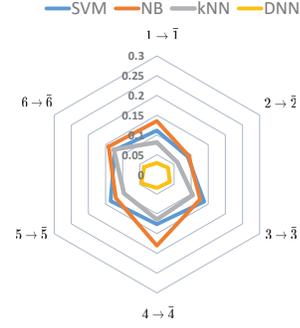}
    \caption{Misclassification rate for single antenna selection}\label{radar}
\end{figure}

Actually, TAS is equivalent to a classification system with ML and DL algorithms. In this subsection, we present the misclassification rate for single antenna selection by using the web representation, $\mathrm{SNR}=15~\mathrm{dB}$.
The value of each point in polygon denotes the misclassification rate of the corresponding channel index by $l \to \overline l $, where $l, \overline l \in \mathcal{L}$ and $l \ne \overline l $.
From Fig.~\ref{radar}, it can be seen that the DL misclassification rate is greatly lower than the rate for DL, which shows the high classification accuracy and great decoupling capability of the DNN by another aspect.

\section{conclusions}\label{conclusion}

In this paper, we applied DNN-based antenna selection scheme to achieve decoupling not being solved by other ML-based schemes (i.e., SVM, NB and k-NN) as well as to reduce the complexity in untrusted relay networks. The simulation results show that our proposed DNN-based scheme can achieve almost the same secrecy rate and SOP as conventional scheme with transmitted power constraint at the relay.


\ifCLASSOPTIONcaptionsoff
  \newpage
\fi

\bibliographystyle{IEEEtran}
\bibliography{ref}

\begin{thebibliography}{10}
\providecommand{\url}[1]{#1}
\csname url@samestyle\endcsname
\providecommand{\newblock}{\relax}
\providecommand{\bibinfo}[2]{#2}
\providecommand{\BIBentrySTDinterwordspacing}{\spaceskip=0pt\relax}
\providecommand{\BIBentryALTinterwordstretchfactor}{4}
\providecommand{\BIBentryALTinterwordspacing}{\spaceskip=\fontdimen2\font plus
\BIBentryALTinterwordstretchfactor\fontdimen3\font minus
  \fontdimen4\font\relax}
\providecommand{\BIBforeignlanguage}[2]{{%
\expandafter\ifx\csname l@#1\endcsname\relax
\typeout{** WARNING: IEEEtran.bst: No hyphenation pattern has been}%
\typeout{** loaded for the language `#1'. Using the pattern for}%
\typeout{** the default language instead.}%
\else
\language=\csname l@#1\endcsname
\fi
#2}}
\providecommand{\BIBdecl}{\relax}
\BIBdecl

\bibitem{Li2017Intelligent}
R.~Li, Z.~Zhao, Z.~Xuan, G.~Ding, C.~Yan, Z.~Wang, and H.~Zhang, ``Intelligent
  {5G}: When cellular networks meet artificial intelligence,'' \emph{IEEE
  Wireless Communications}, vol.~PP, no.~99, pp. 2--10, 2017.

\bibitem{D2017Deep}
S.~Dorner, S.~Cammerer, J.~Hoydis, and S.~T. Brink, ``Deep learning-based
  communication over the air,'' \emph{IEEE Journal of Selected Topics in Signal
  Processing}, vol.~PP, no.~99, pp. 1--1, 2017.

\bibitem{Oshea2017An}
T.~Oshea, J.~Hoydis, T.~Oshea, and J.~Hoydis, ``An introduction to deep
  learning for the physical layer,'' \emph{IEEE Transactions on Cognitive
  Communications \& Networking}, vol.~3, no.~4, pp. 563--575, 2017.

\bibitem{He2018Transmit}
D.~He, C.~Liu, T.~Q.~S. Quek, and H.~Wang, ``Transmit antenna selection in
  {MIMO} wiretap channels: A machine learning approach,'' \emph{IEEE Wireless
  Communications Letters}, vol.~PP, no.~99, pp. 1--1, 2018.

\bibitem{8419776}
W.~Lee, ``Resource allocation for multi-channel underlay cognitive radio
  network based on deep neural network,'' \emph{IEEE Communications Letters},
  vol.~22, no.~9, pp. 1942--1945, Sep. 2018.

\bibitem{7852251}
E.~Nachmani, Y.~Be'ery, and D.~Burshtein, ``Learning to decode linear codes
  using deep learning,'' in \emph{2016 54th Annual Allerton Conference on
  Communication, Control, and Computing (Allerton)}, Sep. 2016, pp. 341--346.

\bibitem{O2017Deep}
T.~J. O'Shea, T.~Erpek, and T.~C. Clancy, ``Deep learning based {MIMO}
  communications.'' \emph{arXiv preprint arXiv:1707.07980}, 2017.

\bibitem{icc_yao}
R.~Yao, Y.~Zhang, N.~Qi, and T.~A. Tsiftsis, ``Machine learning-based antenna
  selection in untrusted relay networks. available:
  https://arxiv.org/abs/1812.10318, preprint,'' \emph{arXiv preprint
  arXiv:1812.10318}, 2018.

\bibitem{yao_access}
R.~Yao, Y.~Lu, T.~A. Tsiftsis, N.~Qi, T.~Mekkawy, and F.~Xu, ``Secrecy
  rate-optimum energy splitting for an untrusted and energy harvesting relay
  network,'' \emph{IEEE Access}, vol.~6, pp. 19\,238--19\,246, 2018.

\bibitem{yao_tifs}
T.~Mekkawy, R.~Yao, T.~A. Tsiftsis, F.~Xu, and Y.~Lu, ``Joint beamforming
  alignment with suboptimal power allocation for a two-way untrusted relay
  network,'' \emph{IEEE Transactions on Information Forensics and Security},
  vol.~13, no.~10, pp. 2464--2474, 2018.

\bibitem{Liu2017A}
W.~Liu, Z.~Wang, X.~Liu, N.~Zeng, Y.~Liu, and F.~E. Alsaadi, ``A survey of deep
  neural network architectures and their applications ¡î,''
  \emph{Neurocomputing}, vol. 234, pp. 11--26, 2017.

\bibitem{Joung2016Machine}
J.~Joung, ``Machine learning-based antenna selection in wireless
  communications,'' \emph{IEEE Communications Letters}, vol.~20, no.~11, pp.
  2241--2244, 2016.

\end{thebibliography}

\end{document}